\documentclass[aps,prc,preprint,superscriptaddress,groupedaddress,showpacs,floatfix,]{revtex4}
\usepackage{graphicx}

\begin{document}

\title{Connecting scaling with short-range correlations}
\author{D.~Berardo}
\affiliation{Dipartimento di Fisica Teorica, Universit\`{a} di Torino,\\
Via P. Giuria 1, 10125 Torino, Italy}
\author{M.~B.~Barbaro}
\affiliation{Dipartimento di Fisica Teorica, Universit\`{a} di Torino and INFN,\\
Sezione di Torino, Via P. Giuria 1, 10125 Torino, Italy}
\author{R.~Cenni}
\affiliation{INFN, Sezione di Genova, Via Dodecaneso 33, I-16146 Genova, Italy}
\author{T.~W.~Donnelly}
\affiliation{Center for Theoretical Physics, Laboratory for Nuclear Science and
Department of Physics, Massachusetts Institute of Technology, Cambridge,
Massachusetts 02139, USA}
\author{A.~Molinari}
\affiliation{Dipartimento di Fisica Teorica, Universit\`{a} di Torino and INFN,\\
Sezione di Torino, Via P. Giuria 1, 10125 Torino, Italy}


\date{\today}


\begin{abstract}
We reexamine several issues related to the physics of scaling in
electron scattering from nuclei. A basic model is presented in which
an assumed form for the momentum distribution having both long- and
short-range contributions is incorporated in the single-particle
Green function. From this one can obtain saturation of nuclear
matter for an NN interaction with medium-range attraction and
short-range repulsion, and can obtain the density-density
polarization propagator and hence the electromagnetic response and
scaling function. For the latter, the shape of the scaling function
and how it approaches scaling as a function of momentum transfer are
both explored.
\end{abstract}

\maketitle

\section{Introduction}

\label{sec:intro}

Scaling phenomena as realized in electroweak interactions with
nuclei at intermediate-to-high energies have several facets,
including scaling of the first kind~\cite{Day:1990mf} (independence
of the momentum transfer $q$) and second
kind~\cite{Donnelly:1998xg,Donnelly:1999sw} (independence of nuclear
species, for instance, as characterized by the Fermi momentum
$k_{F}$). In addition, universality for different reaction channels
(longitudinal, transverse, \textit{etc.}) has been called scaling of
the zeroth kind~\cite{Donnelly:1999sw,Antonov:2009qg}, while
universality in the isoscalar and isovector channels has been called
scaling of the third kind \cite{Caballero:2007tz}. All four types of
scaling are reasonably well respected by data at sufficiently high
energies, namely away from threshold and for momentum transfers
typically twice the Fermi momentum or larger, although there are
observed to be scaling violations and their origins provide
interesting insights into the dynamics of the scattering processes.
For example, the transverse EM response is known to involve both the
familiar one-body currents and two-body Meson-Exchange Currents
(MEC) which do not scale in the same way
\cite{Amaro:2002mj,DePace:2003xu,DePace:2004cr,Amaro:2009dd,Amaro:2010iu}.

In the present study we focus on one particular aspect of scaling,
namely, scaling of the first kind. Our motivation is to explore the
interconnections between the strong interaction dynamics of a
representative NN potential that is chosen to provide the correct
binding energy and saturation density of nuclear matter, on the one
hand, and a corresponding Green function that is made to be
consistent with those properties of nuclear matter, on the other.
Having such a Green function one can immediately obtain the
density-density polarization propagator and hence the longitudinal
response $R_L$ and scaling function $F_L$. For brevity these are
called simply $R$ and $F$ in the rest of this work. In contrast to
the usual approach, in the present study we ``work backwards"
assuming a form for the momentum distribution $n(k)$ and, given
this, obtaining the corresponding energy per particle as a
consequence. In particular, the chosen momentum distribution is
taken to have both long-range contributions (those below and
slightly above the Fermi surface) and short-range contributions
which give rise to a tail that extends to high momentum.

To be able to carry out this study with much of the development
still analytic we restrict our attention to the non-relativistic
situation and assume a translationally invariant (infinite,
homogeneous) many-body system of point nucleons. Our goal is not to
provide a detailed numerical study of scaling phenomena for
comparison with experiment, as this is better done with relativistic
modeling, but is to provide insight into how the short-range part of
the momentum distribution which is important for obtaining
saturation of nuclear matter also has a role to play in the
corresponding scaling phenomena. We shall, however, explore several
properties of scaling and scaling violations that are observed
experimentally, namely, how scaling is approached for large momentum
transfers $q$ (in the scaling region it is approached from above,
something most models fail to explain) and whether or not the
present model is capable of explaining the observed asymmetry found
in the scaling function.

In passing let us draw some comparisons with Deep Inelastic
Scattering (DIS) of leptons on the proton which has been understood
on the basis of the Bjorken scaling law and the parton model. There
also one observes scaling violations which in the high-energy
situation are coped with using the so-called evolution equations.
Two basic assumptions lie at the foundations of Bjorken scaling
(see, for example~\cite{parisi}): (1) the highly virtual photon
interacts with the proton through point-like constituents (the
partons) and (2) the partons cannot change their momenta during the
extremely short time interval available for the DIS process and the
parton-parton interactions are very weak, a situation referred to as
asymptotic freedom. When taken into account the latter leads to
scaling violations. In contrast, the situation of interest in the
present work on electron scattering from strongly interacting
nucleons in nuclei is apparently quite different. The model we use
does not display the equivalent of asymptotic freedom and yet
scaling is quite well obeyed, despite the strength of the partonic
(nucleonic here) interactions. Of course our model will not be able
to account for all types of scaling violations, namely those which
stem from partonic sub-structures (gluons in QCD versus mesons in
nuclei via meson exchange currents; the latter have been the subject
in other studies, for instance, \cite{DePace:2003xu,DePace:2004cr}).
In other words our nucleons (partons) are viewed as point-like. The
very strong correlations between the nucleons induced by the
short-range repulsion appear, at least in our model, to lend
themselves to a description in terms of a mean field framework, in
which the nucleons do not interact, apart from Pauli correlations.
In other words the effect of the hard core is embedded in the
modification of the nucleon momentum distribution with respect to
that of the Fermi gas, still keeping an independent-particle model
for the system. Thus to the extent that the mean field provides a
realistic description of nuclei at large momenta, then scaling
should occur, as we have found.

This paper, which is closely connected to the research developed
in~\cite{Barbaro:2009iv}, outlines the model in
Sect.~\ref{sec:model}. Section~\ref{sec:manybody} addresses the
problems of linking the model to conventional perturbation theory
and focuses on the Coulomb Sum Rule (CSR), a quantity crucially
dependent upon the pair correlation function (pcf) or, equivalently,
upon the momentum distribution $n(k)$ of the nucleons in the
nucleus. Indeed the CSR represents one of the best testing grounds
for the pcf and $n(k)$. In Sect.~\ref{sec:propagator} the fermion
propagator, the key element in constructing the response of our
system to an external probe and hence the scaling function, is set
up. All the issues connected with scaling and the results we have
obtained with our approach are collected in Sect.~\ref{sec:results},
and then finally in Sect.~\ref{sec:concl} we summarize our main
conclusions from this study.

\section{The Model}

\label{sec:model}

The basic formula we start with reads~\cite{Gottfried} (we use $\hbar =c=1$)
\begin{equation}
\frac{E}{A}=\frac{4V}{A}\int \frac{d\vec{k}}{(2\pi )^{3}}\frac{k^{2}}{2m}%
n(k)+\frac{1}{2A}\int d\vec{r}_{1}d\vec{r}_{2}v(\vec{r}_{1}-\vec{r}_{2})C(%
\vec{r}_{1}-\vec{r}_{2})~,  \label{eq:gott}
\end{equation}%
where the factor of $4$ accounts for the spin-isospin degeneracy (a
summation over these variables is of course understood in the second
term as well) and $A$ is the particle number.
Equation~(\ref{eq:gott}) yields the ground-state energy of the
system (actually the energy per particle). We apply it to an
infinite, homogeneous, non-relativistic ensemble of nucleons, viewed
as enclosed in a large volume $V$ to be let to go to infinity at the
end of the calculation. We assume that a two-body force acts between
the nucleons and that this is described by the potential $v(r)$,
where $r=|\vec{r}_{1}-\vec{r}_{2}|$. In Eq.~(\ref{eq:gott})
$C(\vec{r}_{1}-\vec{r}_{2})$ is the pcf simply related by a Fourier
transform to the $n(k)$ in our spatially homogeneous system.

Now, rather than attempting to compute $n(k)$ starting from the
potential $v(r)$ adopting one of the various many-body techniques
available for the purpose (for example a perturbative one), we
\emph{assume} the momentum distribution to be parametrized as
\begin{equation}
n(k)=\theta (k_{F}-k)\Big(1-\alpha \frac{k^{2}}{k_{F}^{2}}\Big)+\theta
(k-k_{F})\beta _{1}e^{-\beta _{2}(\frac{k}{k_{F}}-1)},  \label{eq:nk}
\end{equation}%
accounting both for the existence of a high-momentum tail in $n(k)$,
as suggested by the presently available experimental information, as
well as standard theory (see, for example, a recent treatment of the
momentum distribution and spectral function in \cite{Anton11}), and
for the Luttinger theorem which guarantees the existence of a Fermi
surface for a ``normal" interacting Fermi system. Of course $n(k)$
(Eq.~\ref{eq:nk}) should fulfill the constraint
\begin{eqnarray}
\label{eq:nknorm} \frac{A}{V} &=&  4 \int \frac{d \vec{k}}{(2
\pi)^3} n(k) = \frac{2}{\pi^2} \Bigg \{ \int_0^{k_F} k^2 dk \Big(1-
\alpha \frac{k^2}{k_F^2} \Big) + \beta_1 \int_{k_F}^{\infty} k^2 dk
e^{-\beta_2 (\frac{k}{k_F} -1)} \Bigg \} \nonumber \\ &=& \frac{2
k_F^3}{3 \pi^2} \bigg ( 1 -\alpha \frac{3}{5} +
3\frac{\beta_1}{\beta_2^3}  (\beta_2^2 +2 \beta_2+2) \bigg ) =
\frac{2 k_F^3}{3 \pi^2} h(\alpha, \beta_1, \beta_2 ) = n_0,
\end{eqnarray}
$n_0$ being the system's constant density.

>From Eq.~(\ref{eq:nk}) the pcf is obtained according to the
definition
\begin{eqnarray}
\label{eq:pcf0}  C(\vec{r}_1 - \vec{r}_2) &=& \sum_{\gamma,\delta}
<\Psi_0|\hat{\Psi}_{\gamma}^{\dagger}(\vec {r}_1)
\hat{\Psi}_{\delta}^{\dagger}(\vec {r}_2) \hat{\Psi}_{\delta}(\vec
{r}_2) \hat{\Psi}_{\gamma} (\vec {r}_1)|\Psi_0> \nonumber \\ &=&
\bigg ( \frac{A}{V} \bigg )^2 - 4 \bigg [ h(\alpha, \beta_1, \beta_2
) \int \frac{d \vec{k}}{(2 \pi)^3} e^{- i \vec{k}\cdot \vec{r}} n(k)
\bigg ]^2 =n_0^2 \Big \{ 1 - \frac{1}{4}g^2 (r) \Big \}
\end{eqnarray}
and one gets
\begin{eqnarray}
\label{eq:pcf}  C(r)  && = n_0^2 \Bigg \{ 1 - \frac{1}{4} \bigg \{
\frac{3}{k_F r} \bigg [ j_1(k_F r) -  \frac{\alpha}{(k_F r)^4} \Big
( 3 ((k_F r)^2 -2) \sin(k_F r)-
\nonumber\\
&& - k_F r ((k_F r)^2 -6) \cos(k_F r) \Big ) + \beta_1 \Big( \frac{k_F r} {(k_F r)^2 +\beta_2^2} \Big )^2 \times \\
&& \times \Big ( \sin (k_F r) \Big ( \beta_2 + \frac{\beta_2^2+
\beta_2^3}{(k_F r)^2} -1 \Big ) +\cos (k_F r) \Big ( k_F r +\frac{2
\beta_2 +  \beta_2^2}{k_F r} \Big ) \Big )\bigg ]
 \bigg \}^2 \Bigg \} \nonumber  ~.
\end{eqnarray}
In Eq.~(\ref{eq:pcf0}) $|\Psi _{0}>$ is the system's ground state
and $\hat{\Psi}_{\gamma }(\vec{r}_{1})$ and $\hat{\Psi}_{\delta
}(\vec{r}_{2})$ are the fermion fields. Furthermore the above
formula holds valid for infinite nuclear matter. If we need the pcf
only for protons, as in the case of the CSR, then the factor
$\frac{1}{4}$ should clearly be replaced with a factor $\frac{1}{2}$
in front of the second term on the right-hand side of
Eq.~(\ref{eq:pcf}) and one should set $n_{0}=Z/V$, $Z$ being the
number of protons. As seen in the above equations the pcf naturally
splits into a direct and an exchange contribution (the first and the
second terms on the right-hand side of Eq.~(\ref{eq:pcf0})).

In passing we note that by setting $\alpha =\beta _{1}=\beta _{2}=0$
in Eq.~(\ref{eq:pcf}), namely by recovering the theta-function
momentum distribution of a non-interacting Fermi system, we get back
the well-known pcf of a Fermi gas
\begin{equation}
C(r)=n_{0}^{2}\bigg
[1-\frac{1}{4}\bigg(\frac{3j_{1}(k_{F}r)}{k_{F}r}\bigg)^{2}\bigg]~
\label{eq:num4}
\end{equation}%
which identifies $g(r)$ with $\frac{3j_{1}(k_{F}r)}{k_{F}r}$.

Now by inserting Eqs.~(\ref{eq:nk}) and (\ref{eq:pcf}) into
Eq.~(\ref{eq:gott}) we obtain the binding energy per particle of the
nuclear system providing the potential $v(r)$ is known. For the
latter we employ a schematic model which retains only the basic
features of the nucleon-nucleon force since, as already emphasized,
our aim is not a precise reproduction of the experimental data.
Accordingly we employ a mixture of a Wigner and a Majorana force,
namely
\begin{equation}
v(r)=u(r)[1-\gamma +\gamma P_{x}]=\left\{
\begin{array}{ccc}
+U_{0} & \gamma =0 & r\leq a \\
-V_{0} & \gamma =\frac{1}{2} & a\leq r\leq b \\
0 &  & b\leq r ,
\end{array}%
\right.  , \label{eq:num5}
\end{equation}%
where $P_{x}$ is the space exchange operator and $\gamma $ a parameter
varying in the range $0\leq \gamma \leq 1$. In accord with common wisdom we
choose for $\gamma $ the values indicated in Eq.~(\ref{eq:num5}) where one
recognizes the strong short-range repulsion and the moderate
intermediate-range attraction characterizing the nucleon-nucleon interaction.

Now all of the elements to compute the behaviour of the $E/A$ versus $k_{F}$
are available. We get
\begin{eqnarray}
\frac{E}{A}=\frac{4}{n_{0}}\frac{4\pi}{(2 \pi)^3}
\frac{1}{2m}\int_{0}^{\infty }dkk^{4}n(k) &+&\frac{n_{0}}{2}4\pi
\int_{0}^{\infty }drr^{2}\Big [U_{0}\theta
(a-r)-\frac{3}{8}V_{0}\theta (b-r)\theta (r-a)\Big ]  \label{eq:BE} \\
&-&\frac{n_{0}}{2}4\pi \frac{1}{4}\int_{0}^{\infty }drr^{2}\Big [U_{0}\theta
(a-r)+\frac{3}{2}V_{0}\theta (b-r)\theta (r-a)\Big ]g^{2}(r)~.
\nonumber
\end{eqnarray}%
The numerical factors appearing in the potential terms stem from
summing over the spin-isospin variables of the interacting nucleons;
they of course enter differently in the direct and exchange
contributions to $E/A$, namely the second and the third terms on the
right-hand side of Eq.~(\ref{eq:BE}).

Before actually displaying the behaviour of $E/A$ versus $k_{F}$ we
have to face the crucial problem of fixing the values of the seven
parameters (four for the potential $v(r)$, three for the momentum
distribution $n(k)$) needed to render our approach predictive. In
order to tackle this problem we proceed as follows: we start by
choosing \textquotedblleft reasonable\textquotedblright\ values.
Next we compute Eq.~(\ref{eq:BE}) using these chosen values and
repeat the procedure adjusting at each step the parameters until
they yield $E/A$ $\sim $ -16 MeV and, for the compression modulus,
$\kappa =\frac{k_{F}^{2}}{9}\frac{\partial ^{2}E/A}{\partial
k_{F}^{2}}\sim $ 14 MeV~\cite{ACM} at the minimum of the curve, that
should occur at a value of $k_{F}$ which, when inserted into
Eq.~(\ref{eq:nknorm}), provides the experimental density of nuclear
matter, namely $0.17 $ fm$^{-3}$.

The three above mentioned constrains (energy, density and compressibility)
turn out to be fulfilled by choosing
\begin{equation}
\alpha =0.2,\ \ \ \beta _{1}=0.4,\ \ \ \beta _{2}=4  \label{eq:num6}
\end{equation}%
and
\begin{equation}
U_{0}=2.5\text{ GeV},\ \ \ V_{0}=53\text{ MeV},\ \ \ a=0.465\text{
fm},\ \ \ b=2.10\text{ fm}~. \label{eq:num6a}
\end{equation}
With these values the minimum of the curve yielding $E/A$ versus $k_{F}$
occurs at $k_{F}$=1.23 fm$^{-1}$, which is obviously different from the
value of the pure Fermi gas, namely $k_{F}$=1.36 fm$^{-1}$ .

\begin{figure}[tbph]
\centering
\includegraphics[height=7cm]{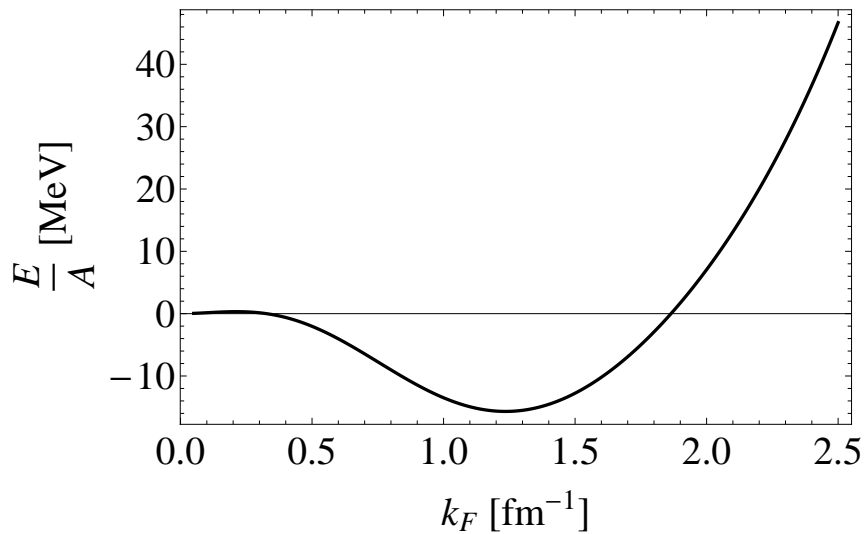}
\caption{The binding energy ($E/A$) versus $k_{F}$, as given by Eq.~(\protect
\ref{eq:BE}). }
\label{fig:BE}
\end{figure}

The associated $E/A$ versus $k_{F}$ is displayed in
Fig.~\ref{fig:BE} which yields
\begin{equation}
\label{eq:BEris} \left( \frac{E}{A}\right)_{\text{min}} = - 15.68
\text{ MeV} , \ \ \ (k_F)_{\text{min}} = 1.23    \text{ fm}^{-1} , \
\ \ (\kappa)_{\text{min}}  = 13.8 \text{ MeV} ~,
\end{equation}
namely the experimental values.

\begin{figure}[th]
\centering
\includegraphics[height=7cm]{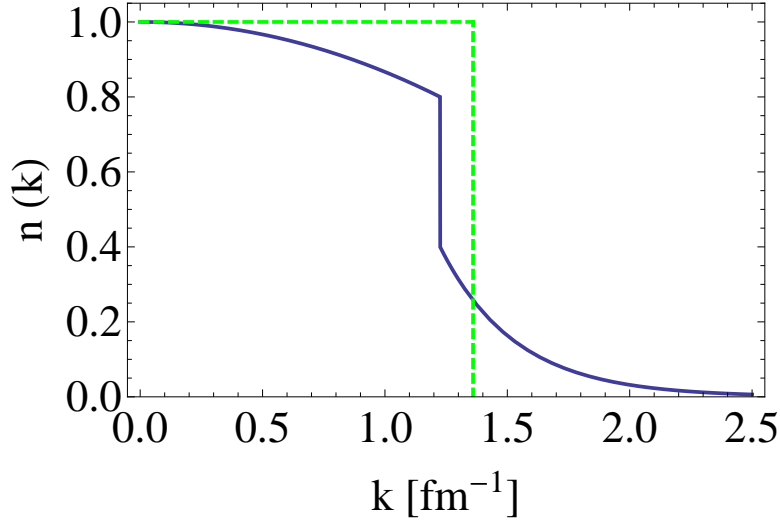}
\caption{(color online) Solid (blue online): the momentum distribution $n(k)$
versus $k$. Also displayed (dashed, green online) is the Fermi gas
$n(k)=\protect\theta (k_{F}-k)$ with $k_{F}=1.36$ fm$^{-1}$.}
\label{fig:nk}
\end{figure}

We also display the momentum distribution $n(k)$ in
Fig.~\ref{fig:nk}. Concerning the results of this section, it should
be clear that the choice of the parameters given in
Eq.~(\ref{eq:num6a}) is far from unique, and that these values
should be viewed as providing a first orientation on a complex
problem. Our goal is only to use something representative for the
potential in attempting to shed light on the connections between
saturation of nuclear matter and electron scattering scaling
phenomena.

\section{The many-body content of $n(k)$ and the CSR}

\label{sec:manybody}

Next the question arises: which are the Feynman diagrams one has to
take into account in order to obtain the $n(k)$ given by
Eq.~(\ref{eq:nk})? This is equivalent to asking: what kind of
correlations among nucleons are responsible for changing the
theta-function into our $n(k)$ which we assumed to be the true
momentum distribution of our system?

\begin{figure}[tbph]
\centering
\includegraphics[height=5cm]{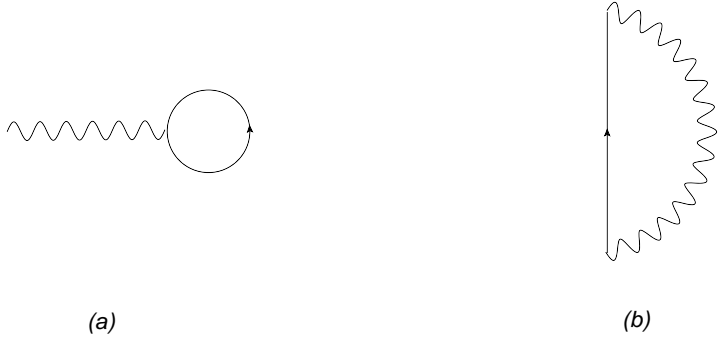}
\caption{First-order proper self-energy $\Sigma _{(1)}^{\star }$; in (a) and
(b) are shown the direct and exchange terms, respectively.}
\label{fig:diagr}
\end{figure}

\begin{figure}[tbph]
\centering
\includegraphics[height=7cm]{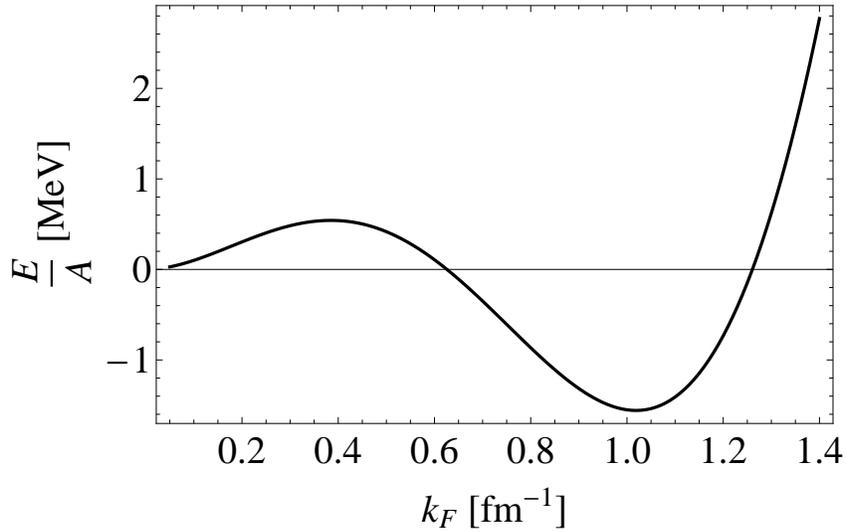}
\caption{$E/A$ versus $k_{F}$ in HF theory employing the potential
of Eq.~ (\protect\ref{eq:num5}) with the parameters of
Eq.~(\protect\ref{eq:num6a}). } \label{fig:HF}
\end{figure}

To help in better grasping the relevance of this question it is of
importance to realize that if we replace in Eq.~(\ref{eq:gott})
$n(k)$ with $\theta (k_{F}-k)$ and $C(\vec{r}_{1}-\vec{r}_{2})$ with
Eq.~(\ref{eq:num4}) we obtain
\begin{eqnarray}
\frac{E}{A} &=&\frac{3}{5}\frac{k_{F}^{2}}{2m}+n_{0}\bigg [\frac{2}{3}\pi
a^{3}U_{0}-\frac{\pi }{4}(b^{3}-a^{3})V_{0}\bigg ]  \nonumber \\
&&-\frac{1}{\pi }\bigg (U_{0}-\frac{3}{2}V_{0}\bigg )
\bigg [\text{Si}(2k_{F}a)+\frac{\cos (2k_{F}a)-3}{2k_{F}a}+\frac{\cos (2k_{F}a)-1}{%
2(k_{F}a)^{3}}+\frac{\sin (2k_{F}a)}{(k_{F}a)^{2}}\bigg ]  \nonumber \\
&&-\frac{3}{2\pi }V_{0}\bigg [\text{Si}(2k_{F}b)+\frac{\cos (2k_{F}b)-3}{%
2k_{F}b}+\frac{\cos (2k_{F}b)-1}{2(k_{F}b)^{3}}+\frac{\sin (2k_{F}b)}{%
(k_{F}b)^{2}}\bigg ]~,  \label{eq:num7}
\end{eqnarray}%
namely the result provided by Hartree-Fock (HF) theory~\cite{ACM}, which, as
is well known, captures the content of the two first-order perturbative
diagrams shown in Fig.~\ref{fig:diagr}.
It is instructive to look at the result one gets in HF with the
parameters given in Eq.~(\ref{eq:num6a}) for the double square well
potential of Eq.~(\ref{eq:num5}). This is displayed in
Fig.~\ref{fig:HF}: here we see that $E/A$ in HF still saturates,
although at a far too low density ($k_{F}=1.02$ fm$^{-1}$) and with
a far too low energy ($E/A=-1.56$ MeV).

To prove that by summing all perturbative diagrams one would recover
the $n(k)$ of Eq.~(\ref{eq:nk}) (and hence the $E/A$ of
Fig.~\ref{fig:BE}) is clearly an impossibility. In the present study
our approach is only to explore a simple model while attempting to
maintain as high a level of coherence as we can. To achieve the
latter we first resort to Calogero's theorem~\cite{Amado:1976zz},
which states that for a homogeneous infinite system asympotically
one should have
\begin{equation}
n(k)\sim \bigg [2mn_{0}\frac{\tilde{v}(k)}{k^{2}}\bigg ]^{2} ,
\label{eq:cal}
\end{equation}%
where, in the present case,

\begin{equation}
\tilde{v}(k)=\int d\vec{r}e^{-i\vec{k}\cdot \vec{r}}v(r)=\frac{4}{3}\pi a^{3}
\bigg \{U_{0}\frac{3j_{1}(ka)}{ka}-V_{0}\bigg [\frac{b^{3}}{a^{3}}\frac{%
3j_{1}(kb)}{kb}-\frac{3j_{1}(ka)}{ka}\bigg ]\bigg \}~.  \label{eq:pottras}
\end{equation}

\begin{figure}[tbph]
\centering
\includegraphics[height=7cm]{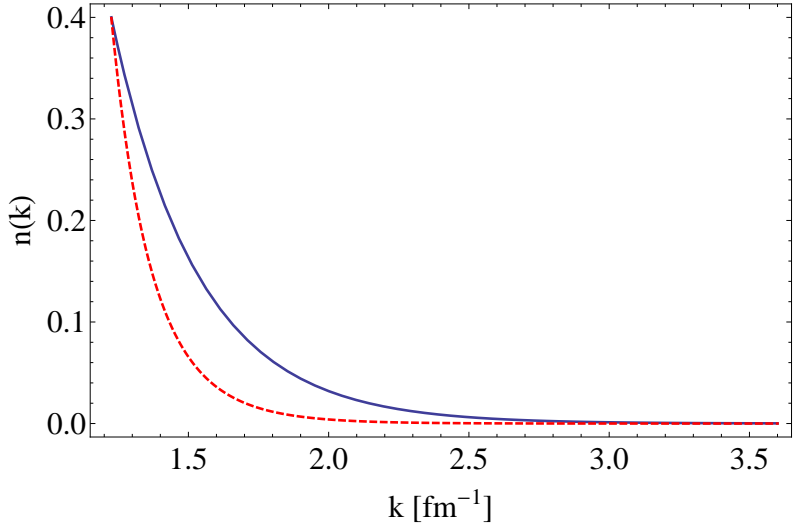}
\caption{(color online) Solid line (blue online): the momentum distribution of
Eq.~(\protect\ref{eq:nk}). Dashed line (red online): Calogero asymptotic
behaviour Eq.~(\protect\ref{eq:cal}), multiplied by a factor
$2\times 10^{-4}$ (see text for the origin of this factor).}
\label{fig:cal}
\end{figure}

In Fig.~\ref{fig:cal} we compare our $n(k)$ with the Calogero's
theorem predictions: we do so in the range of momenta starting from
the Fermi surface ($k=k_{F}=1.23$ fm$^{-1}$) up to $k\sim 3.6$
fm$^{-1}$, since for larger momenta both distributions become so
small that they render the comparison meaningless. In
Fig.~\ref{fig:cal} one sees that the two curves are not too
different; hence our $n(k)$ and $\widetilde{v}(r)$, while not
exactly the same, nevertheless display an acceptable degree of
coherence. It should be stated that in carrying out such a
comparison, unfortunately, Calogero's theorem does not quantify the
value of $q$ signalling the onset of the asymptotic regime, and
hence we have arbitrarily normalized the Calogero's asymptotic
momentum distribution in such a way to have it coincide with our
$n(k)$ at the Fermi surface.

Concerning the question related to the measurement of $n(k)$, we
recall that access to information on this is offered by the CSR,
which in fact essentially depends \textquotedblleft
only\textquotedblright\ upon the momentum distribution at least
within the context of the present model.
Indeed from unitarity one has
\begin{equation}
S(q)=Z-n_{0}^{2}V\frac{1}{2}\int d\vec{r}e^{-i\vec{q}\cdot
\vec{r}}g^{2}(r), \label{eq:num8}
\end{equation}%
where $g(r)$ (see Eq.~(\ref{eq:pcf})) is directly fixed by the
Fourier transform of our $n(k)$.

\begin{figure}[tbph]
\centering
\includegraphics[height=7cm]{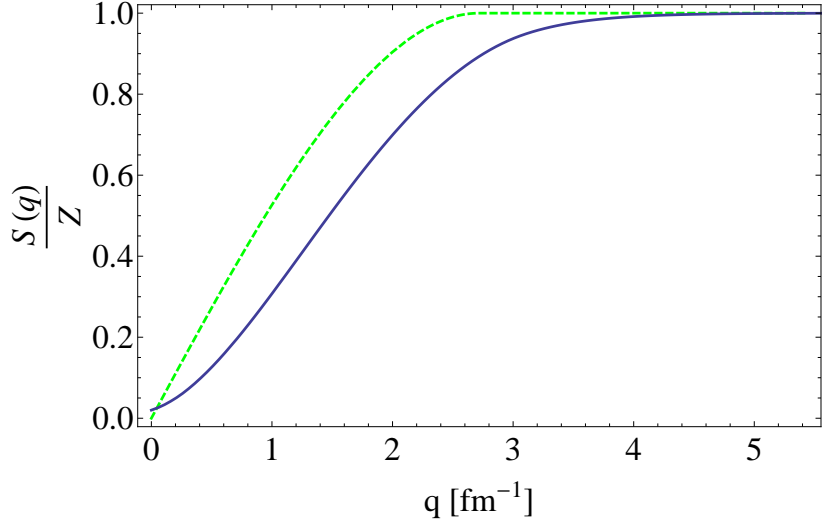}
\caption{(color online) The CSR for a free Fermi gas (dashed line, green online)
and for a correlated one according to our model (solid line, blue online).}
\label{fig:CSR}
\end{figure}

For sake of completness we display in Fig.~\ref{fig:CSR} the CSR
predicted by our $n(k)$. Actually this curve was already shown
in~\cite{Barbaro:2009iv}, however we revisit it once more here to
illustrate how the attainement of the asymptotic value (namely one)
is postponed by the nucleon-nucleon correlations, in particular, the
repulsive short-range ones, to larger values of $q$ ($\cong 4.6$
fm$^{-1}$) than for the Fermi gas situation ($\cong 2.46$
fm$^{-1}$).

To illustrate our second path toward coherence we commence by
recalling the alternative model of \cite{Caballero:2010fn}, which
deals with the same issues treated in the present paper, namely the
scaling function and momentum distribution, but with a different
philosophy. The study of \cite{Caballero:2010fn} is based on the use
of PWIA (Plane-Wave Impulse Approximation) and has as its starting
point the assumption of the factorization of the one-particle
exclusive cross section (namely the cross section for the process
$(e,e^{\prime }N)$) according to the expression
\begin{equation}
\bigg [\frac{d\sigma }{d\epsilon ^{\prime }d\Omega ^{\prime }dp_{N}d\Omega
_{N}}\bigg ]^{\text{PWIA}}=K\sigma ^{eN}(q,\omega ;p,\mathcal{E},\phi
_{N})S(p,\mathcal{E})~.  \label{eq:num9}
\end{equation}%
In the above $K$ is a kinematical factor, $\sigma ^{eN}$ the $eN$
single-nucleon cross section and $S(p,\mathcal{E})$ the nucleon spectral
function expressed in terms of the so-called missing-energy and
missing-momentum variables (for their definition see, for
instance,~\cite{Cenni:1996zh}).
We do not dwell here on the procedure leading from Eq.~(\ref%
{eq:num9}) to the scaling function: this topic has been addressed a number
of times in the literature. Suffice it to say that the key step is the
integration of Eq.~(\ref{eq:num9}) over the variables $p$ and $\mathcal{E}$
in a domain which has been analyzed in the past and whose boundaries are set
by energy and momentum conservation. The above procedure corresponds to
passing from the semi-inclusive ($e,e^{\prime }N$) cross section to the
inclusive ($e,e^{\prime }$) cross section. Indeed one has
\begin{equation}
\frac{d\sigma }{d\epsilon ^{\prime }d\Omega ^{\prime }}=\overline{\sigma }%
^{eN}(q,\omega ;p=|y|,\mathcal{E}=0)F(q,\omega ),  \label{eq:num10}
\end{equation}%
where $\overline{\sigma }^{eN}$ is evaluated at the minimum values of $%
\mathcal{E}$ and $p$ allowed by kinematics. The scaling function
$F(q,\omega )$ is easily found to read
\begin{equation}
F(q,\omega )=2\pi \int \int_{\Sigma (q,\omega )}pdpd\mathcal{E}S(p,\mathcal{E%
})  \label{eq:num11}
\end{equation}%
in PWIA. Now if the integration domain $\Sigma (q,\omega )$ in Eq.~(\ref%
{eq:num11}) extends up to encompass the limiting value $\mathcal{E}%
\rightarrow \infty $ then the scaling function can be linked to the momentum
distribution according to~\footnote{Note that a different normalization
for the momentum distribution is adopted in
Refs.~\cite{Cenni:1996zh,Caballero:2010fn}.}
\begin{equation}
n(p)=\int_{0}^{\infty }d\mathcal{E}S(p,\mathcal{E})\,.  \label{eq:num12}
\end{equation}

Thus, at variance with our method which is based on the coherence
between the momentum distribution and the scaling function, in the
sense of having both of them derived within the same theory, in
\cite{Caballero:2010fn} the strategy is to exploit the link between
the scaling function and the momentum distribution and hence the
possibility of extracting information on $n(k)$ from ($e,e^{\prime
}$) experimental data.

\section{Setting up the single-nucleon propagator}

\label{sec:propagator}

Our scheme is based on the central role played by the single-fermion
propagator $G(\vec{k},\omega )$ for nucleons in the nucleus. In
fact, as is well known, knowledge of the latter gives access to the
$E/A$ according to the expression \cite{FW71}
\begin{equation}
E=\frac{iV}{2(2\pi )^{4}}\lim_{\eta \rightarrow 0^{+}}\int d\vec{k}%
\int_{-\infty }^{\infty }d\omega e^{i\omega \eta }\bigg (\frac{k^{2}}{2m}%
+\omega \bigg )\text{Tr}G(\vec{k},\omega )  \label{eq:num13}
\end{equation}%
and to the scaling function according to the formula
\begin{equation}
F(q,\omega )=-\frac{q}{m}
\frac{V}{\pi }\text{Im}\Pi (q,\omega )~.  \label{eq:num14}
\end{equation}%
This represents the content of linear response theory and, as in \cite%
{Caballero:2010fn}, assumes the factorization of the single-nucleon
cross section. In Eq.~(\ref{eq:num14}) $\Pi (q,\omega )$ is referred
to as the polarization propagator or, in coordinate space, as the
density-density correlation function. In field theory language it
corresponds to a particular choice of the field arguments in the
two-particle propagator and in momentum space reads
\begin{equation}
\Pi (q,\omega )=-\frac{i}{2}\int \frac{d^{4}k}{(2\pi )^{4}}G(k+q)G(k).
\label{eq:num15}
\end{equation}%
Knowing $G$ ensures control of $\Pi (q,\omega )$ and this, in turn, through
Eq.~(\ref{eq:num14}), permits the determination of the scaling function $%
F(q,\omega )$. Also, in a coherent scheme, $G(k)$ should yield the correct
momentum distribution (namely our $n(k)$).

\begin{figure}[tbph]
\centering
\includegraphics[height=7cm]{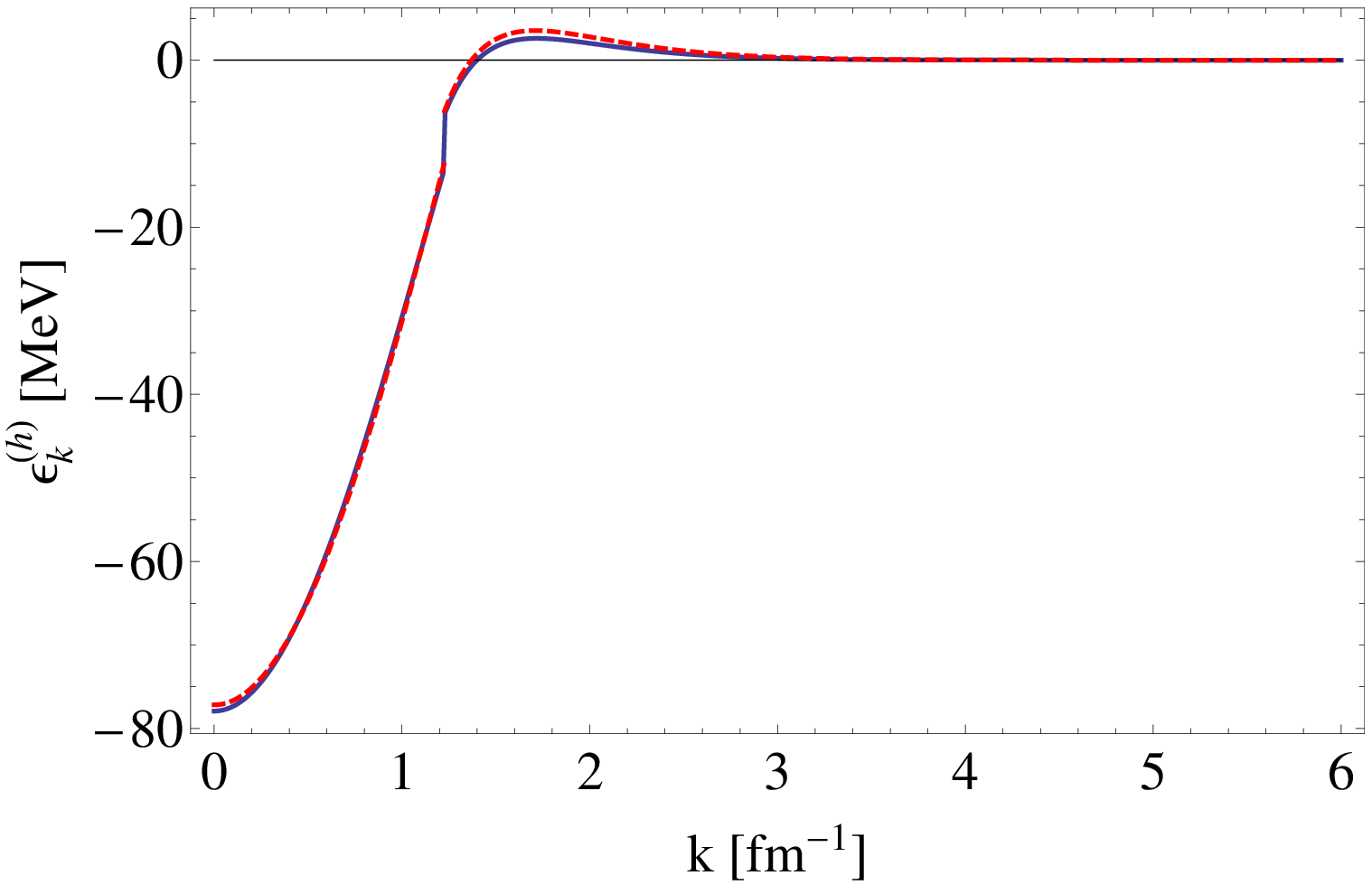}
\caption{(color online) Single-particle energy (hole) of our model shown for $k_{F}=1.23$ fm%
$^{-1}$ which corresponds to the saturation density of the system.
The dashed line represents the fit given by Eq.~(\ref{eq:fit1}). }
\label{fig:omegabuco}
\end{figure}

\begin{figure}[tbph]
\centering
\includegraphics[height=7cm]{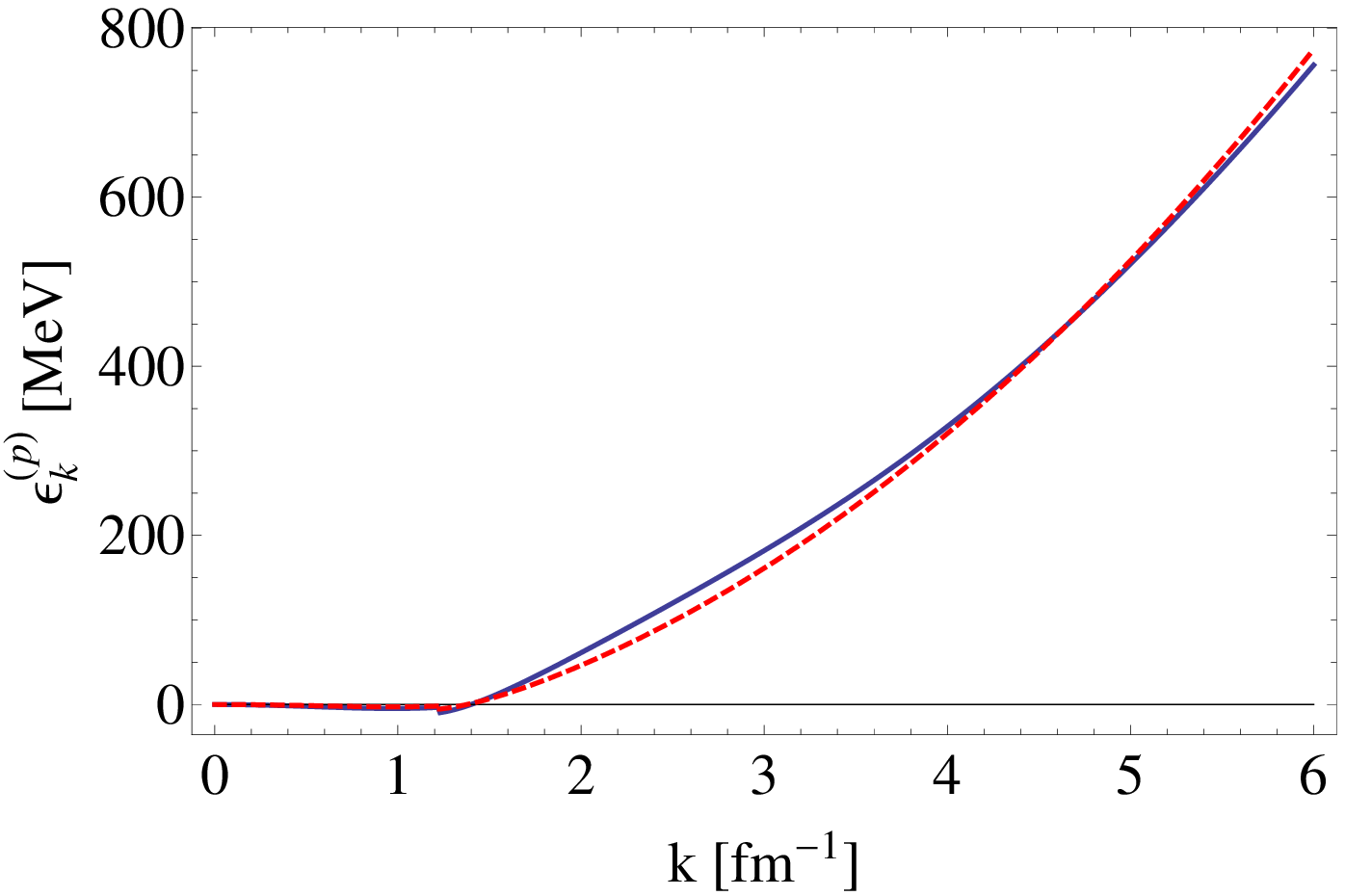}
\caption{(color online) Single-particle energy (particle) of our
model shown for $k_{F}=1.23 $ fm$^{-1}$ which corresponds to the
saturation density of the system. The dashed line represents the fit
given by Eq.~(\ref{eq:fit2}). } \label{fig:omegapart}
\end{figure}

Can such a propagator be derived on the basis of the knowledge of
$n(k)$ alone? The answer is yes in the simple approach taken in the
present work where an infinite, homogeneous many-body system of
nucleons has been assumed. This kind of mean-field approximation
possesses a remarkably coherent structure; whether this remains true
in a more sophisticated many-body framework has yet to be proven.
Continuing to work within the context of our simple model, as a
first step to achieve this goal we rewrite our basic expression in
Eq.~(\ref{eq:gott}), exploiting the Faltung theorem of Fourier
analysis. For this purpose, we start from
\begin{eqnarray}
\frac{E}{A} &=&\frac{4V}{A}\int \frac{d\vec{k}}{(2\pi )^{3}}\frac{k^{2}}{2m}%
n(k)+\frac{n_{0}}{2}\int d\vec{r}v(r)\bigg (1-\frac{1}{4}g^{2}(r)\bigg )
\nonumber \\
&=&\frac{4}{n_{0}}\int \frac{d\vec{k}}{(2\pi )^{3}}\frac{k^{2}}{2m}n(k)+%
\frac{n_{0}}{2}\bigg [\int d\vec{r}v_{D}(r)-\frac{1}{4}\int d\vec{r}%
v_{E}(r)g^{2}(r)\bigg ]  \label{eq:BE2}
\end{eqnarray}%
with $g^{2}(r)$ given by Eq.~(\ref{eq:pcf0}) and the direct and
the exchange potential terms by Eq.~(\ref{eq:BE}) in coordinate
space and introducing the Fourier representations according to
\begin{equation}
\tilde{v}_{D}(k)=\int d\vec{r}e^{-i\vec{k}\cdot \vec{r}}v_{D}(r)=\frac{4\pi
}{3}U_{0}a^{3}\frac{3j_{1}(ka)}{ka}-\frac{\pi }{2}V_{0}\bigg (b^{3}\frac{%
3j_{1}(kb)}{kb}-a^{3}\frac{3j_{1}(ka)}{ka}\bigg)~,  \label{eq:vtild}
\end{equation}
\begin{equation}
\tilde{v}_{E}(k)=\int d\vec{r}e^{-i\vec{k}\cdot \vec{r}}v_{E}(r)=\frac{4\pi
}{3}\bigg [a^{3}\frac{3j_{1}(ka)}{ka}\bigg (U_{0}-\frac{3}{2}V_{0}\bigg )+%
\frac{3}{2}V_{0}b^{3}\frac{3j_{1}(kb)}{kb}\bigg ]  \label{eq:vtile}
\end{equation}%
and
\begin{equation}
\int d\vec{r}e^{-i\vec{k}\cdot \vec{r}}\bigg (1-\frac{1}{4}g^{2}(r)\bigg)%
=(2\pi )^{3}\delta (\vec{k})-\frac{4}{n_{0}^{2}}h^2(\alpha, \beta_1,
\beta_2 )\int \frac{d\vec{p}}{(2\pi )^{3}}n(p)n(|\vec{k}-\vec{p}|).
\label{eq:num16}
\end{equation}%
Then, with the help of Eqs.~(\ref{eq:vtild}--\ref{eq:num16}), we obtain
\begin{equation}
\frac{E}{A}=\frac{4}{n_{0}}\int \frac{d\vec{k}}{(2\pi )^{3}}n(k)\bigg [\frac{%
k^{2}}{2m}+\frac{n_{0}}{2}\tilde{v}_{D}(0)-\frac{h^2(\alpha,
\beta_1, \beta_2
)}{2}\int \frac{d\vec{q}}{(2\pi )^{3}}%
n(q)\tilde{v}_{E}(|\vec{k}+\vec{q}|)\bigg ]=\frac{4}{n_{0}}\int \frac{d\vec{k%
}}{(2\pi )^{3}}\epsilon _{k}^{(h)},  \label{eq:num17}
\end{equation}%
where
\begin{equation}
\epsilon _{k}^{(h)}=n(k)\bigg [\frac{k^{2}}{2m}+\frac{n_{0}}{2}\tilde{v}%
_{D}(0)-\frac{h^2(\alpha, \beta_1, \beta_2
)}{2(2\pi )^{2}}\int_{0}^{\infty }dpp^{2}\tilde{v}%
_{E}(p)\int_{-1}^{1}dxn(|\vec{p}-\vec{k}|)\bigg ]  \label{eq:num18}
\end{equation}%
is the single-hole energy displayed in Fig.~\ref{fig:omegabuco}. For
comparison, in Fig.~\ref{fig:omegapart} the single-particle energy
\begin{equation}
\epsilon _{k}^{(p)}=\frac{1-n(k)}{n(k)}\epsilon _{k}^{(h)}  \label{eq:num19}
\end{equation}%
is shown. Note the discontinuity of $\sim 6.5$ MeV in both $\epsilon
_{k}^{(h)} $ and $\epsilon _{k}^{(p)}$ at the Fermi surface, the
vanishing at large $k$ of $\epsilon _{k}^{(h)}$ and the value
\begin{eqnarray}
\epsilon _{0}^{(h)}
&=&\frac{n_{0}}{2}\tilde{v}_{D}(0)-\frac{h^2(\alpha, \beta_1,
\beta_2
)}{(2\pi)^{2}}%
\int_{0}^{\infty }dpp^{2}\tilde{v}_{E}(p)n(p)  \nonumber \\
&=&\frac{n_{0}}{2}\frac{4\pi }{3}a^{3}\bigg [U_{0}-\frac{3}{8}V_{0}\bigg (%
\frac{b^{3}}{a^{3}}-1\bigg )\bigg ]  \nonumber \\
&&-\frac{h^2(\alpha, \beta_1, \beta_2
)}{\pi }\Bigg \{\bigg (U_{0}-\frac{3}{2}V_{0}\bigg )\bigg [\text{Si}%
(k_{F}a)-(1-\alpha )\sin (k_{F}a)-3\alpha j_{1}(k_{F}a)\bigg ]  \nonumber \\
&&\qquad \qquad \quad \quad \quad \quad \,\, \,\,\,\quad +\frac{3}{2}V_{0}\bigg [\text{Si}%
(k_{F}b)-(1-\alpha )\sin (k_{F}b)-3\alpha j_{1}(k_{F}b)\bigg ]  \nonumber \\
&&+\beta _{1}\bigg (U_{0}-\frac{3}{2}V_{0}\bigg )\bigg [k_{F}a\frac{%
k_{F}a\sin (k_{F}a)-\beta _{2}\cos (k_{F}a)}{(k_{F}a)^{2}+\beta _{2}^{2}}%
+e^{\beta _{2}}\text{Im}\text{E}_{1}(\beta _{2}-ik_{F}a)\bigg]  \nonumber \\
&&\quad \qquad \quad +\beta _{1}\frac{3}{2}V_{0}\bigg [k_{F}b\frac{%
k_{F}b\sin (k_{F}b)-\beta _{2}\cos (k_{F}b)}{(k_{F}b)^{2}+\beta _{2}^{2}}%
+e^{\beta _{2}}\text{Im}\text{E}_{1}(\beta _{2}-ik_{F}b)\bigg]\Bigg \}
\nonumber \\
&=&-77.90\text{ MeV}  \label{eq:num20}
\end{eqnarray}%
of the latter at the origin.

The above single-particle energies yield the poles of the fermion
propagator which, as a consequence, can then be cast into the form
\begin{eqnarray}
G(k,\omega ) &=&\frac{n(k)}{\omega -n(k)\big [\frac{k^{2}}{2m}+\frac{n_{0}}{2%
}\tilde{v}_{D}(0)-\frac{h^2(\alpha, \beta_1, \beta_2
)}{2(2\pi )^{2}}\int_{0}^{\infty }dpp^{2}\tilde{v}%
_{E}(p)\int_{-1}^{1}dxn(|\vec{p}-\vec{k}|)\big ]-i\eta }  \label{eq:G} \\
&&+\frac{1-n(k)}{\omega -[1-n(k)]\big [\frac{k^{2}}{2m}+\frac{n_{0}}{2}%
\tilde{v}_{D}(0)-\frac{h^2(\alpha, \beta_1, \beta_2
)}{2(2\pi )^{2}}\int_{0}^{\infty }dpp^{2}\tilde{v}%
_{E}(p)\int_{-1}^{1}dxn(|\vec{p}-\vec{k}|)\big ]+i\eta }~. \nonumber
\end{eqnarray}%
This structure of the propagator tells us that in our model the
holes exist below, but also above, the Fermi surface. Likewise the
particles exist above, but also below, the Fermi surface. These
occurrences clearly reflect the behaviour of our momentum
distribution. The important point to be stressed, however, is that
the propagator in Eq.~(\ref{eq:G}) provides the correct system
energy and $n(k)$.

Now, to pave the way to the actual evaluation of $\Pi $, it helps to
realize that, although the expression of the single-particle
energies is far from being simple (in fact it cannot be expressed
analytically), nevertheless its $k$ dependence lends itself to be
suitably represented, apart from the factor $n(k)$, by a parabola.
This is reminiscent of HF theory. Hence we use the following quite
faithful representation
\begin{eqnarray}
\label{eq:fit1} \epsilon^{(h)}(k) = n(k) (A_{(h)} + B_{(h)} k^2) =
n(k) \Big (A_{(h)} +  \frac{k^2}{2 m_{(h)}^{\star}} \Big )
\end{eqnarray}
with $A_{(h)}= -77.16$ MeV and $B_{(h)}= 41.10$ MeV fm$^{2}$,
yielding an effective mass $m_{(h)}^{\star} = 0.50$ $m$ for holes
and
\begin{eqnarray}
\label{eq:fit2} \epsilon^{(p)}(k) = (1-n(k) ) (A_{(p)} + B_{(p)}
k^2) = (1 - n(k) ) \Big (A_{(p)} +  \frac{k^2}{2 m_{(p)}^{\star}}
\Big )
\end{eqnarray}
with $A_{(p)} = -43.09$ MeV and $B_{(p)}= 22.72$ MeV fm$^{2}$,
yielding an effective mass $m_{(p)}^{\star} = 0.91$ $m$ for
particles. It is indeed startling to realize how large the impact of
our two-body interaction in Eq.~(\ref{eq:num5}) is on the effective
hole mass, the generally accepted ratio being actually $m^*/m\simeq
0.83$~\cite{Johnson:1987zza}. How faithful Eq.~(\ref{eq:fit1}) and
Eq.~(\ref{eq:fit2}) are in providing the single-particle energies,
can be garnered from Figs.~\ref{fig:omegabuco} and
\ref{fig:omegapart}.

Finally, the response function of the system is easily derived,
\begin{equation}
R(q,\omega )=-\frac{V}{\pi }\text{Im}\Pi (q,\omega ),
\label{eq:num23}
\end{equation}%
and from the response one immediately obtains the scaling function per
proton according to
\begin{eqnarray}
F(q,\omega ) &=&\frac{q}{m}\frac{R(q,\omega )}{Z}  \label{eq:scalf2} \\
&=&\frac{q}{m}\frac{1}{n_{0}}\frac{1}{\pi ^{2}}\int_{0}^{\infty
}dkk^{2}n(k)\int_{-1}^{1}dx[1-n(|\vec{k}+\vec{q}|)]\delta \lbrack
\omega -\epsilon ^{(p)}(|\vec{k}+\vec{q}|)+\epsilon ^{(h)}(k)] ,
\nonumber
\end{eqnarray}%
where the trivial frequency and azimuthal integrations have been
performed. The results of our numerical calculations of Eq.
(\ref{eq:scalf2}) are reported in the next section. Following
standard practice when discussing scaling we shall show results for
the dimensionless scaling function
\begin{equation}
f(q,\omega)\equiv k_F \times F(q, \omega) \label{eq:littlef}
\end{equation}
which takes on an especially simple form for the relativistic Fermi
gas (see, {\it e.g.,} \cite{Donnelly:1998xg,Donnelly:1999sw}).

\section{Results}
\label{sec:results}

\begin{figure}[htbp]
\centering
\includegraphics[height=7cm]{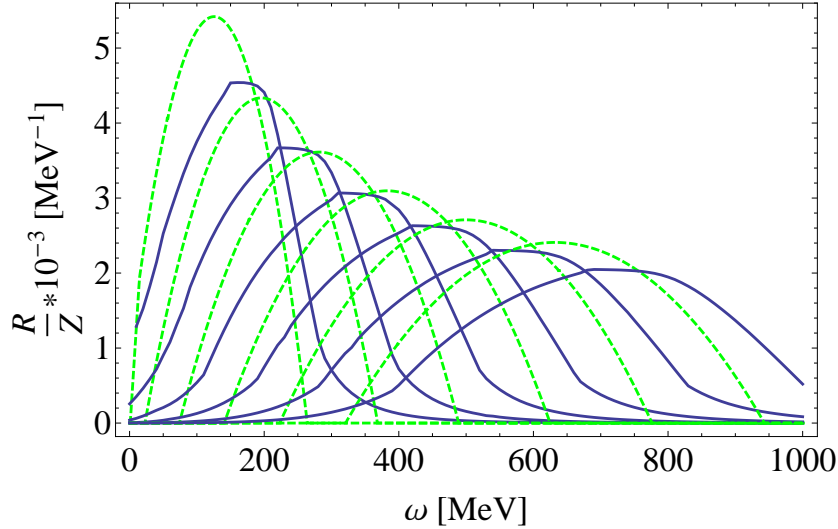}
\caption{(color online) The response function of our model (solid
line, blue online) and the response function of a free Fermi gas (dashed line,
 green online)
plotted versus $\omega$ for $q = 2 k_F = 2.46$ fm$^{-1}$ up to $q =
4.5 k_F = 5.53$ fm$^{-1}$ in steps of $0.5 k_F$. Results for lower
values of $q$ peak at lower values of $\omega$.}
\label{fig:rispvsomega}
\end{figure}

\begin{figure}[htbp]
\centering
\includegraphics[height=7cm]{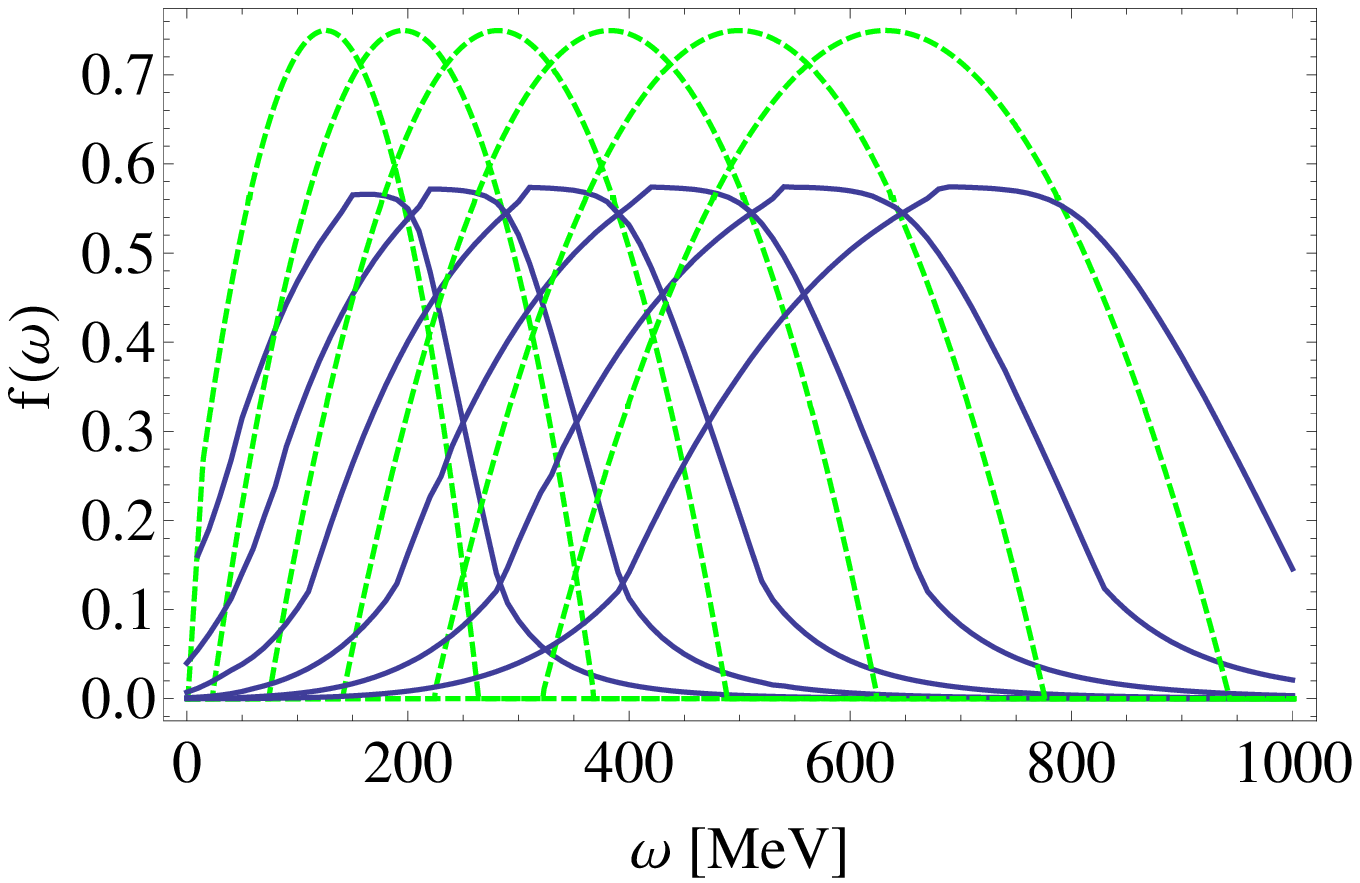}
\caption{(color online) The scaling function of our model (solid
line, blue online) and the scaling function of a free Fermi gas (dashed line,
green online)
plotted versus $\protect\omega$ for the same values of $q$ used in
Fig. \ref{fig:rispvsomega}. } \label{fig:scalvsomega}
\end{figure}

In Figs.~\ref{fig:rispvsomega} and \ref{fig:scalvsomega} the results
for the response $R(q,\omega )$ (Eq.~(\ref{eq:num23})) and scaling
function $f(q,\omega )$ (Eq.~(\ref{eq:littlef})) are shown versus
$\omega $ for a range of momentum transfers. The response and the
scaling function for the free Fermi gas are also displayed for
comparison.

We observe the following:

\begin{itemize}

\item The response and scaling function obtained using our
model, as expected, span a range of energy loss that extends to
larger values than that seen for the Fermi gas model --- a clear
indication of the role of correlations among the nucleons. The
widths seen in our model are somewhat larger than those of the Fermi
gas and the peak heights are somewhat lower, both in better accord
with experimental data.

\item The peak positions in our model are shifted to higher energy
loss than for the Fermi gas (see also the discussions to follow).

\item Unlike for the Fermi gas model our $R(q,\omega )$ and $f(q,\omega )$ are
no longer perfectly symmetric around their maxima. While
approximately so, they have tails that extend both to higher and
lower values of $\omega$. However, the degree of asymmetry is not as
large as what is observed experimentally.

\item Note that while the height of $R(q,\omega )$ decreases with $q$,
the height of $f(q,\omega )$ remains constant.

\end{itemize}

\begin{figure}[tbph]
\centering
\includegraphics[height=7cm]{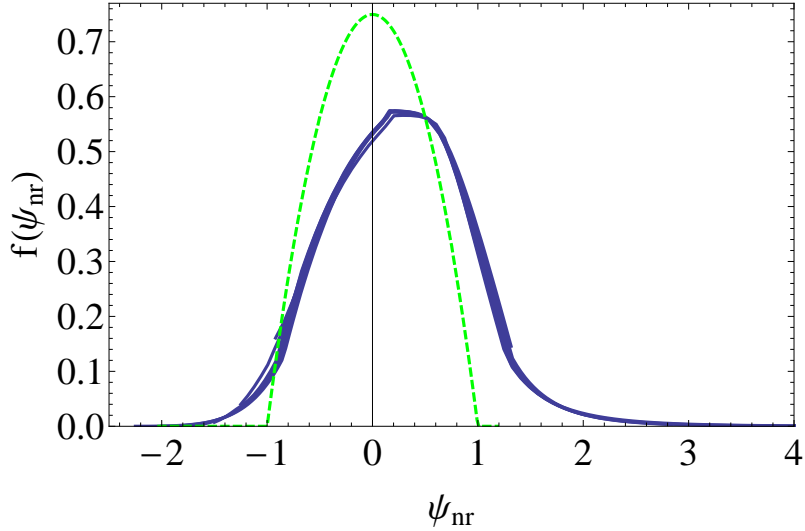}
\caption{(color online) The scaling function of our model (blue online)
displayed versus the non-relativistic variable $\protect\psi _{nr}$. The
momentum transfer range is the same as in the previous two figures.
For reference the Fermi gas result is shown as a dashed curve (green online);
clearly, by construction, it scales perfectly. } \label{fig:scalvsY}
\end{figure}

\begin{figure}[tbph]
\centering
\includegraphics[height=7cm]{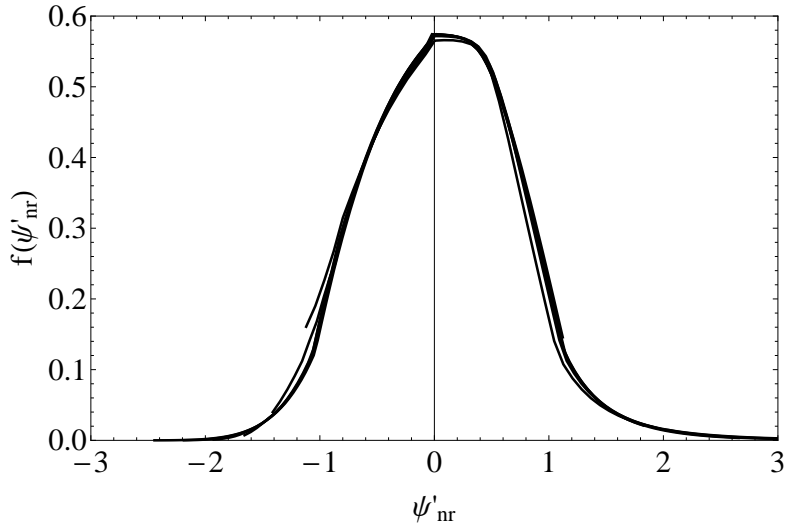}
\caption{The scaling function of our model displayed
versus the scaling variable of Eq.~(\protect\ref{eq:num25}) for the
same values of $q$ used in Fig. \ref{fig:scalvsY}. }
\label{fig:scalvsYprime}
\end{figure}

\begin{figure}[tbph]
\centering
\includegraphics[height=7cm]{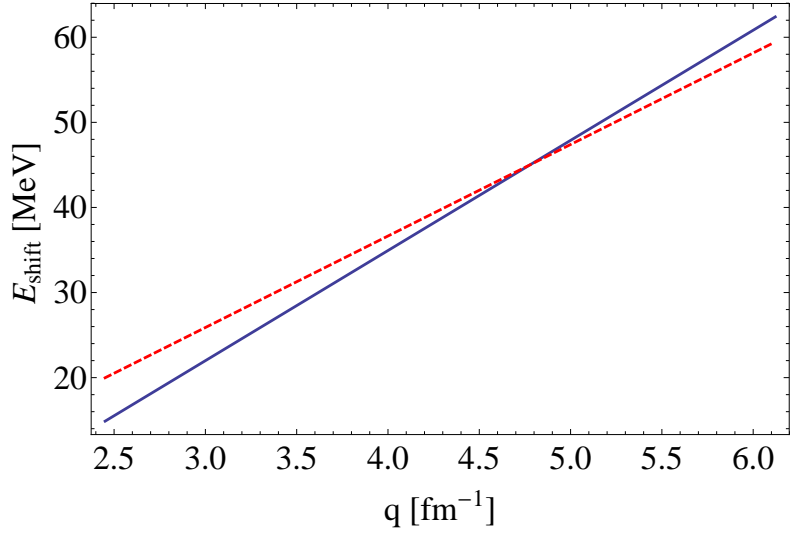}
\caption{(color online) The $q$-dependent energy shift in
Eq.~(\protect\ref{eq:eshift}) (solid curve, blue online)
together with the energy shift obtained in RMF studies of $^{12}$C
(dashed curve, red online). }
\label{fig:shiftvsq}
\end{figure}

To investigate the scaling behaviour of our results we follow the
usual procedures and display $f$, not versus $\omega $ as above, but
versus the well-known non-relativistic scaling
variable~\cite{Alberico:1988bv}
\begin{equation}
\psi _{nr}=\frac{1}{k_{F}}\bigg (\frac{m\omega }{q}-\frac{q}{2}\bigg
). \label{eq:num24}
\end{equation}%
Now we indeed see in Fig.~\ref{fig:scalvsY} that the scaling
functions for different values of $q$ tend to group together very
closely when displayed versus $\psi _{nr}$, that is they scale.
Noting that the coalescence in our model occurs at a peak value
other than $\psi _{nr}=0$, it is interesting to investigate whether
a simple modification of the scaling variable different from the one
of Eq.~(\ref{eq:num24}) can be devised to shift the peak position to
zero. One can always do this by employing the variable
\begin{equation}
\psi _{nr}^{\prime }\equiv \frac{1}{k_{F}}\bigg (\frac{m\omega ^{\prime }}{q}-%
\frac{q}{2}\bigg ),  \label{eq:num25}
\end{equation}%
where $\omega ^{\prime }=\omega -E_{shift}(q)$, and where
$E_{shift}(q)$ is a $q$-dependent energy shift. If one uses the
simple parametrization
\begin{equation}
E_{shift}(q)= E_0+E_1 (q/k_F) \label{eq:eshift} \end{equation} with
$E_0 = -17.4$ MeV and $E_1 = 15.9$ MeV (see
Fig.~\ref{fig:shiftvsq}), then almost perfect scaling centered about
$\psi _{nr}^{\prime }=0$ is attained, as seen in
Fig.~\ref{fig:scalvsYprime}. It appears that the impact of the NN
interactions that have been incorporated in the present model is
felt via a $q$-dependent shift in the definition of this new scaling
variable, although a direct connection to the underlying dynamics is
not obvious. As is clear from Fig.~\ref{fig:shiftvsq} similar
shifting is seen using relativistic mean field theory. One of course
should not expect these to be identical, since the present model is
non-relativistic while the RMF results are obtained using a
relativistic model. In either case, while the shift is qualitatively
what is observed experimentally, it is probably somewhat too strong
in both cases~\cite{Caballero:2006wi,Amaro:2006if}.

\begin{figure}[tbph]
\centering
\includegraphics[height=7cm]{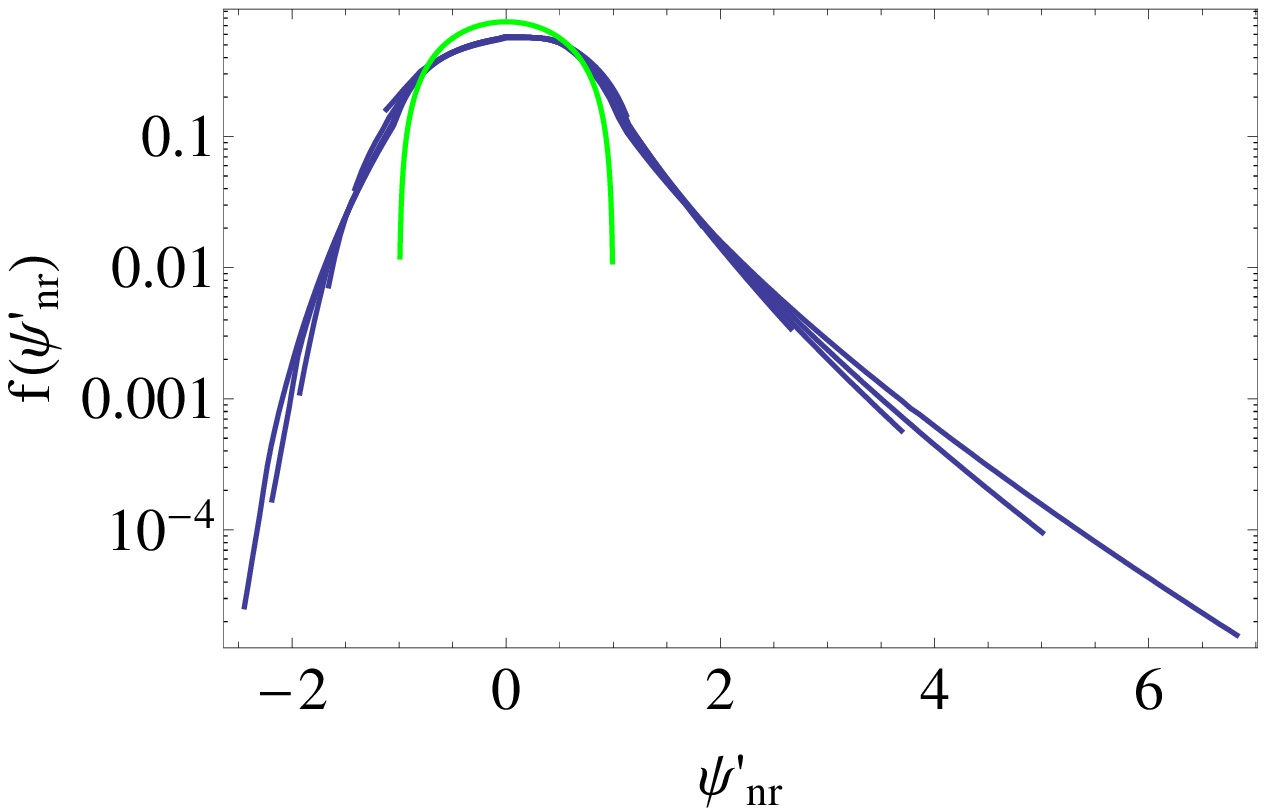}
\caption{(color online) As for Fig. \ref{fig:scalvsYprime}, but now
on a semilog scale. } \label{fig:logscalvsY}
\end{figure}

In Fig.~\ref{fig:logscalvsY} we show the same results as in
Fig.~\ref{fig:scalvsYprime}, but now on a semilog scale. The
asymmetry, while small, is clearly apparent. More strength is
shifted to higher values of $\psi_{nr}^\prime$ and, whereas the
Fermi gas cuts off abruptly and is only nonzero within the Fermi
cone, the present model produces strength extending to very large
and small values of $\psi_{nr}^\prime$, in accord with experiment.
It is worth remarking that if the high momentum tail in the momentum
distribution in Eq.~(\ref{eq:nk}) is set to zero ({\it i.e.,}
$\beta_1$ is set to zero) then these tails extending to large
$|\psi_{nr}^\prime|$ essentially disappear. While setting $\beta_1$
to zero is not simply setting the part of the momentum distribution
arising from short-range correlations to zero, since the long-range
correlations also move some strength from below the Fermi surface to
somewhat above it, it is very suggestive that in the present model
the origin of the tails in the scaling function are principally due
to the short-range physics, as is often assumed to be the case.
Interestingly, the position of the peak is largely unaffected by
``turning off" the high-$k$ part of the momentum distribution,
suggesting that the peak position is not strongly correlated with
the short-range physics, but rather with the long-range physics.

\begin{figure}[tbph]
\centering
\includegraphics[height=7cm]{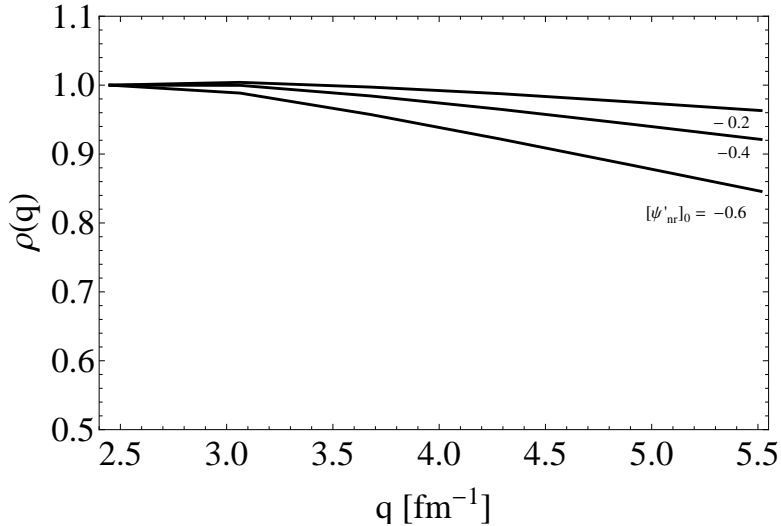}
\caption{The scaling function versus $q$ for
$[\psi^\prime_{nr}]_0=-0.6$, -0.4 and -0.2, {\it i.e.,} in the
scaling region.} \label{fig:approach}
\end{figure}

In concluding this section we address the problem of how the scaling
regime is approached. We do so using the scaling variable
$[\psi^\prime_{nr}]_0$, where, to make this closer to what has been
used in analyses of experimental data \cite{Day:1990mf} a constant
energy shift $E_{shift}=30$ MeV has been chosen in using
Eq.~(\ref{eq:num25}); this is indicated by the subscript ``0". We
display $f$ as a function of $q$ for three values of the scaling
variable in the scaling region, {\it i.e.,} to the left of the QE
peak. The curves are shown normalized at $q=2 k_F$, namely, the
ratio
\begin{equation} \rho (q) \equiv f(q,[\psi^\prime_{nr}]_0)/f(2
k_F,[\psi^\prime_{nr}]_0)
\label{eq:psip0}
\end{equation}
is displayed. From Fig.~\ref{fig:approach} it clearly appears that
the scaling regime is approached from above, an occurrence which is
qualitatively in accord with experimental findings.

\section{Conclusions}
\label{sec:concl}

In the present study we have developed a model centered around an
assumed form for the momentum distribution of nucleons in nuclei.
The momentum distribution has been chosen to reflect current
understanding of how dynamical effects underlying the nuclear
many-body problem lead to a form for $n(k)$ with both low-$k$
components coming from long-range interactions and an extended tail
at high-$k$ arising from short-range interactions. We have
restricted the scope of the study to infinite, homogeneous nuclear
matter, have employed only point nucleons, have assumed a strictly
non-relativistic model (although in future work we hope to extend
the scope to relativistic modeling), and have restricted our
attention to the longitudinal electromagnetic response, for the
present.

Working with this as a basis we have developed the formalism in two
different directions, maintaining as much consistency as possible.
First, we have devised a single-particle Green function that leads
to the know saturation properties of nuclear matter and to realistic
particle and hole single-particle energy spectra. Second, we have
taken the same Green function to obtain the density-density
polarization propagator and, through its imaginary part, have
obtained the longitudinal electron scattering response function
$R(q,\omega)$ and the scaling function $f(q,\omega)$. For the latter
we have explored several aspects of scaling and of scaling
violations.

We find that scaling is quite well respected, despite the strength
of the NN interactions implicit in the problem. There are seen to be
some scaling violations, for instance, those observed as shifts of
the quasielastic peak positions as functions of $q$ --- indeed these
are also seen in relativistic mean field theory. The shape of the
scaling function is observed to be more spread out than is the Fermi
gas scaling function and the former yields a somewhat lower peak
height than the latter, both in rough accord with experiment. Where
it is large the scaling function in our model is somewhat
asymmetric; however, it is not enough so to agree with experiment.
When the tails of the scaling function are examined in detail the
asymmetry is more apparent and, indeed, the strength at both very
large and very small values of the scaling variable is significant,
in accord with experiment. If the high-$k$ contributions to the
momentum distribution are ``turned off" then these tails disappear,
suggesting that their origin lies in the part of the momentum
distribution arising from short-range correlations. The position of
the peak, however, appears to be due to long-range physics.

In summary, the present study has demonstrated that a high level of
consistency can be maintained in simultaneously representing both
the saturation properties of nuclear matter and the scaling
properties of the longitudinal electron scattering response.

\begin{acknowledgments}
We thank Dr. Arturo De Pace for valuable help in solving several
computational problems.
This work is supported in part (T.W.D.) by the U.S. Department
of Energy under cooperative agreement DE-FC02-94ER40818.
\end{acknowledgments}

\end{document}